\documentclass[prl,twocolumn,aps,amssymb,footinbib,showpacs]{revtex4}
\usepackage{amssymb}
\usepackage{amssymb}
\usepackage{graphicx}
\usepackage{amsmath}
\usepackage{times}
\usepackage{color}
\usepackage{subfigure}
\usepackage{setspace}
\usepackage{dcolumn}
\usepackage{bm}
\input epsf

\begin{document}

\title{Noise-resistant entanglement of strongly interacting spin systems}
\author{ A. M. Rey$^{1}$, L. Jiang$^{2}$,  M. Fleischhauer$^{3}$, E. Demler$^{2}$  and M.D.
Lukin$^{1,2}$}  \affiliation{$^{1}$ Institute for Theoretical
Atomic, Molecular and Optical Physics, Cambridge, MA, 02138.}
\affiliation{$^{2}$ Physics Department, Harvard University,
Cambridge, Massachusetts 02138, USA}
 \affiliation{$^{3}$Fachbereich Physik, Technische Universit\"at Kaiserslautern, D-67663
  Kaiserslautern, Germany}
\date{\today }
\begin{abstract}
We  propose and analyze a  scheme that makes use of interactions
between spins to protect certain correlated many-body states from
decoherence. The method exploits the finite energy gap of properly
designed  Hamiltonians to generate a manifold insensitive to local
noise fluctuations. We apply the scheme  to achieve
decoherence-resistant generation of many particle GHZ states and
show that it can improve the sensitivity in precision spectroscopy
with trapped ions. Finally we also show that cold atoms in optical
lattices interacting via short range interactions can be utilized to
engineer the required long range interactions for  a robust
generation of entangled states.
\end {abstract}


\maketitle


Quantum entanglement has recently emerged as an important resource
in quantum information science
 and metrology \cite{Nielsen,Preskill}. For example, entangled
many-particle states can be used to perform long-distance quantum
communication and scalable quantum computation  and to enhance the
spectroscopic sensitivity in quantum-limited measurements. However,
many-particle entangled
 states are difficult to prepare and to maintain since  they are extremely fragile: in practice, noise and decoherence
rapidly collapse  them  into classical statistical mixtures.

In this Letter we propose and analyze a new method for
 the robust generation
of entangled states
 and its protection
against decoherence. Our approach
is based on creating a degenerate many-body  subspace, which can be
isolated from the rest of the Hilbert space
by
properly designed interactions.   We focus specifically on interacting two-level
systems  where the two states are  associated with  spin
sublevels, e.g trapped ions or neutral atoms\cite{Leibfried,Leibfried2,Mandel,Widera,Romalis}. Such systems are
of particular  importance for applications in precision
measurements. Few-particle entangled states have already
been prepared with trapped ions,
 and proof-of-principle experiments demonstrating
  the improvement
of spectroscopic sensitivity have been carried out \cite{Leibfried,Leibfried2}. Our aim is to
devise a technique to increase the  robustness of entangled collective states.
 We  discuss potential applications
of this protected evolution to perform  precision measurements with
trapped ions and to generate many-particle-GHZ-type states in
optical lattices.

Our approach can be best understood by considering a simple example
a multi-spin system with  isotropic ferromagnetic interactions.
These interactions will naturally align the spins. While all of the spins
can be rotated together around an arbitrary axis without  cost of energy, local spin flips are energetically
forbidden. Consequently ferromagnetic interactions  enable the generation of   superpositions (suitable  for example  for
precision spectroscopy) with a substantial suppression of decoherence (See
Fig.1).
\begin{figure}[h]
\addtolength{\belowcaptionskip}{-0.7cm}\addtolength{\abovecaptionskip}{-0.6cm}
\begin{center}
\leavevmode {\includegraphics[width=1.75 in]{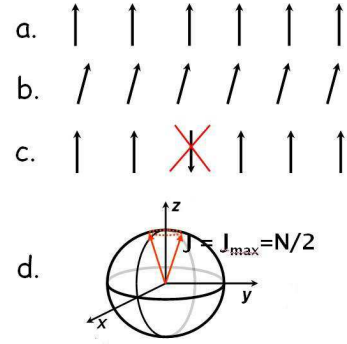}}
\end{center}
\caption{Many-body protected manifold (MPM) generated  by  the
isotropic ferromagnetic exchange:
 (a) The ground manifold is spanned by the states with all spins aligned in the same
 direction,
(b) Any global  rotation about an arbitrary axis is allowed, (c)
local spin flips are energetically suppressed. Panel (d) shows that
the MPM lies on the  Bloch  surface with radius $J=J_{max}=N/2$.}
\end{figure}
We note that in Ref.\cite{Unanyan} gapped quantum systems have  been
proposed for decoherence-free entanglement generation by adiabatic
ground-state transitions. Specifically, Ref.\cite{Unanyan} exploited
the finite energy gap  to protect non-degenerate ground states.
The present mechanism is capable of protecting degenerate
{\it multi-state manifolds}
with a large number of quantum degrees of
freedom and use non-equilibrium dynamics as the mechanism for
entanglement generation.


We consider a collection of
$N$
 spin 1/2 atoms with isotropic infinite
range ferromagnetic interactions described by the Hamiltonian
$\hat{H}=\hat{H}_{prot}+\hat{H}_m $, where
\begin{equation}
\addtolength{\belowdisplayskip}{-0.25
 cm}\addtolength{\abovedisplayskip}{-0.2 cm} \hat{H}_{prot}=-\lambda
\hat{J}^{(0)2} \quad {\rm{and}}\quad \hat{H}_m=\omega_0
 {\hat{J}^{(0)}}_{z}.\label{Hs}
\end{equation}We used ${\hat{J}^{(0)}} _{\alpha }$ to denote
collective spin operators of the $N$ atoms:
${\hat{J}^{(0)}} _{\alpha }=\frac{1}{2}\sum _i\hat{\sigma}_{i}^{\alpha }$, where $\alpha =x,y,z$ and
$\hat{\sigma}_{i}^{\alpha }$ is a   Pauli
 operator acting on the   $i^{th}$ atom.
 Moreover
 we have identified the two relevant internal states of the
atoms with the effective spin index $\sigma
~=~\uparrow,\downarrow$ . These states have energy splitting   $\omega_0$ in
units such that $\hbar =1$.

An appropriate basis to describe the dynamics of the system is the
 basis spanned by collective pseudo-spin  states   denoted as $
|J,M,\beta \rangle _{z}$ \cite{Arecchi72}. These states satisfy the eigenvalue relations $
\hat{J}^{(0)2}|J,M,\beta \rangle _{z}=J(J+1)|J,M,\beta \rangle _{z}$ and $
{\hat{J}^{(0)}}_{z}|J,M,\beta \rangle _{z}=M|J,M,\beta \rangle _{z}$, with $
J=N/2,\dots ,0$ and $-J\leq M\leq J$.  $\beta $ is an additional
quantum number  associated with the permutation group which  is required   to form
a complete set of labels for  all the  $2^{N}$ possible states.

The isotropic Hamiltonian $\hat{H}_{prot}$  has a  ground state manifold spanned by a set of $N+1$ degenerate
states.   They  lie on the surface  of  the  Bloch sphere with
maximal radius $J=N/2$ (see Fig.1) and   are  totally  symmetric, i.e.
invariant with respect to particle  permutations. This set of  states
are fully characterized  by $J$ and  $M$, and   we denote them as $|N/2,M\rangle _{z}$ with the additional label
$\beta $  omitted.  There is a finite energy gap  $E_g=\lambda N$ that isolates the ground state manifold from the
rest of the Hilbert
space. This gap is the key for the many-body protection  against decoherence.
We will refer to the ground state manifold as the \textit{many-body protected manifold} (MPM).

To understand the protection within the MPM,  we first assume
that the dominant type of decoherence is
 single-particle dephasing. Such dephasing comes from  processes that, while preserving
the populations in the atomic levels, randomly change the phases
leading to  a decay of the off-diagonal density matrix
elements. We model the phase
decoherence by adding to Eq. (1) the following Hamiltonian
\cite{Huelga}
\begin{equation}
\addtolength{\belowdisplayskip}{-0.2
cm}\addtolength{\abovedisplayskip}{-0.2cm}
\hat{H}_{env}=\frac{1}{2}\sum_{i}h_{i}(t)\hat{\sigma}_{i}^{z},\label{env}
\end{equation}where the $h_{i}(t)$ are assumed to be independent stochastic  Gaussian
processes with zero mean and with
 autocorrelation function $\overline{h_{i}(t)h_{j}(\tau )}
=\delta _{ij}f(t-\tau )$. Here the bar denotes averaging over the
different random outcomes. In what follows we will use the property that zero mean  Gaussian variables satisfy the
 property $\overline{%
\exp [-i\int_{0}^{t}d\tau h(\tau )]}=\exp [-\Gamma(t)]$, with
$\Gamma(t)
=\frac{1}{2}\int_{0}^{t}dt_{1}\int_{0}^{t}dt_{2}f(t_{1}-t_{2})$.

Let us first study the  $\lambda=0$ dynamics, with  the system  in
the ground manifold at $t=0$. Phase decoherence modifies the free
  precession of the Bloch vector generated by  $\hat{H}_m$. While $\langle{\hat{J}^{(0)}}_{z}(t)\rangle$
 remains a conserved quantity, the $x$ and
$ y$ projections
decay
 exponentially  with    rate  $\Gamma(t)$,
i.e. ${\hat{J}^{(0)}}_{x,y}(t)\rangle = e^{-\Gamma(t)}\langle {\hat{J}^{(0)}}_{x,y}(t)\rangle|_{\Gamma
=0}$,
  where $\langle {\hat{J}^{(0)}}_{x}(t)\rangle|_{\Gamma =0}$ is the expectation
value in the absence of noise \cite{Huelga}. In addition, due to
  local phase fluctuation which
drive the system out of the ground state manifold and  deplete  the $J=N/2$ levels
also
$\langle \hat{J}^{(0)2}\rangle$  decays exponentially.

The effect of the environment  on the free evolution  is significantly reduced by $\hat{H}_{prot}$.
 The protection  can be best  understood by using the basis of
collective states. In terms of collective spin operators $\hat{H}_{env}$ can be written as:
\begin{equation}
\addtolength{\belowdisplayskip}{-0.2
cm}\addtolength{\abovedisplayskip}{-0.2 cm}
\hat{H}_{env}=\frac{1}{\sqrt{N}}\sum_{k=0}^{N-1}g^{k}(t)\hat{J}^{(k)}_z,\label{env2}
\end{equation}where $g^{k}(t)=\frac{1}{\sqrt{N}}\sum_{j=1}^N h_j(t) e^{-i\frac{2 \pi j
k}{N}}$ and
 $\hat{J}^{(k)}_{\alpha}=\frac{1}{2} \sum_{j=1}^N \hat{\sigma}_j e^{i\frac{2 \pi j k}{N}}$.
 In  the presence of a large energy gap $E_g$, one can distinguish two
different types of  processes: (i) Decoherence  effects that take
place  within the MPM
 due to the collective dynamics induced by  the $k=0$ component of  $\hat{H}_{env}$,
  and (ii) transitions   across the gap induced by the inhomogeneous terms
with $k\ne 0$.
The latter couple  the MPM with the
 rest of the system. All allowed  transitions must conserve $M$
since
 both the system and
 noise Hamiltonian commute with $\hat{J}_{z}^{(0)}$.

In the limit when the noise is sufficiently slow (
i.e.
when the spectral density of the noise
 has a cutoff frequency
$\omega_c \ll E_g$)   the effect of phase  decoherence is dramatically
reduced
as type (ii) processes are
energetically forbidden and
  only type (i) processes are effective. In this limit  the noise acts  as
a uniform  random magnetic field.
If at $t=0$
the density matrix,
$\hat{\rho}$, is  $\hat{\rho}=\sum_{M,\tilde{M}}
\rho_{M,\tilde{M}}(0)|\frac{N}{2},M\rangle\langle\frac{N}{2},\tilde{M}|$,
then after time $t$ each component $\rho_{M,\tilde{M}}$  acquires an
additional phase $e^{i (\theta _{M}(t)-\theta _{\tilde{M}}(t))}$
with $\theta _{M}(t)\equiv \frac{M}{\sqrt{N}}\int_{0}^{\tau}g^0
(\tau)$, such that
\begin{equation}
\addtolength{\belowdisplayskip}{-0.2
cm}\addtolength{\abovedisplayskip}{-0.2 cm}
 \overline{
\rho_{M,\tilde{M}}(t)}= \rho_{M,\tilde{M}}(0) e^{i \omega_0
(M-\tilde{M})t} e^{-\Gamma(t)\frac{(M-\tilde{M})^2}{N}}. \label{pde}
\end{equation}Note the factor of $\sqrt{N}$ in the denominator of $\theta_M$. It
is fundamental for the reduction of the effect of decoherence within
the MPM. For example,   ${\hat{J}^{(0)}}_{x,y}$ decays $N$ times
slower than in the unprotected system: i.e.
$\langle{\hat{J}^{(0)}}_{x,y}(t)\rangle=e^{-\Gamma(t)/N}
 \langle{\hat{J}^{(0)}}_{x,y}(t)\rangle|_{\Gamma=0}$.

 Let us  now elaborate more on the necessary conditions required to suppress type (ii) processes.
 For this, we will assume the power spectrum of
the noise, $f(\omega )\equiv \int dte^{-i\omega t}f(t)$, to have a cut-off
frequency $\omega _{c}$ (e.g. $f(\omega )= f$ for $\omega\leq \omega_c$ and 0 otherwise with $f<\lambda$  ).
 A  time dependent
 perturbation analysis predicts
that as long as   $\omega _{c}<E_g$,  the decay rate $\gamma^M(t)$
of  the diagonal elements  $\rho_{M,M}(t)$ of the density matrix due
to type (ii) process is always bounded by $ \overline{\gamma
^{M}(t)}<\left(\frac{N^{2}-4M^{2}}{N^{2}}\right)\left(\frac{f \omega _{c}}{\lambda ^{2}N}%
\right)\ll 1.$ Hence  in the slow noise limit the population of the
MPM is fully preserved.

{\it Protected generation of  N-particle GHZ states:} $N$-particle
Greenberger-Horne-Zeilinger (GHZ)  states \cite{Greenberger} of the form
\begin{equation}
\addtolength{\belowdisplayskip}{-0.2
cm}\addtolength{\abovedisplayskip}{-0.2 cm}
|\psi ^{GHZ}_{x}\rangle \equiv \frac{1}{\sqrt{2}}\left( e^{-i\phi _{+}}|%
\frac{N}{2},\frac{N}{2}\rangle _{x}+e^{i\phi -}|\frac{N}{2},-\frac{N}{2}%
\rangle _{x}\right),\notag \label{ghz}
\end{equation}where $|\frac{N}{2},\pm\frac{N}{2} \rangle _{x}$ are fully polarized
states along the $\pm x$
 and $\phi _{\pm}$ are arbitrary real phases,
 can be generated  by evolving   the state $|N/2,N/2\rangle _{x}$
 with an interaction of the form \cite{Sorensen,Molmer}
\begin{equation}
\addtolength{\belowdisplayskip}{-0.2
cm}\addtolength{\abovedisplayskip}{-0.2 cm}
 \hat{H}_z=\chi
\hat{J}^{(0)2}_{z}.
\end{equation}The generation process can be understood
  by writing  $|N/2,N/2\rangle _{x}$ as $\sum_M C_M|N/2,M\rangle _{z} $.
During the evolution the Hamiltonian
 imprints an $M^2$ dependent phase  to the different  components. As the system evolves,
  at  first the winding of the  phases  leads to a  collapse  of
  $\langle {\hat{J}^{(0)}}_{x}\rangle$. However, at time $\chi t_{rev}=\pi $ all  components  rephase  with
opposite polarization, and a perfect revival of the initial state
 is observed with   $\langle {\hat{J}^{(0)}}_{x}\rangle=-N/2$ (see Fig. 2). Right at
time  $t_0=t_{rev}/2$ the system becomes
an $ N$-particle GHZ state.

Recent experiments
\cite{Leibfried, Leibfried2} have used this type of scheme  to  generate GHZ states in trapped
 ions with the aim to perform precision measurements of
$\omega_0$ ( ideally the use of GHZ states should  enhance the phase
sensitivity to the fundamental Heisenberg limit \cite{Bollinger}).
However, decoherence significantly limited the applicability of the method
and  in practice even for  only six ions, the achieved  phase sensitivity  was
significantly below  the fundamental limit. The effect of decoherence can be quantified by calculating the fidelity of creating a GHZ state
defined as  $\mathcal{F}(t_0)=\overline{\langle \psi
^{GHZ}_x |\hat{\rho}(t_0)|\psi
^{GHZ}_{x}\rangle }$. It is possible to show that, after the evolution with the total Hamiltonian,
$\hat{H}_{env} + \hat{H}_{z} $,
 the fidelity is degraded to:
\begin{equation}
\addtolength{\belowdisplayskip}{-0.25
 cm}\addtolength{\abovedisplayskip}{-0.2 cm}
\mathcal{F}(t_0)=\left(\frac{1+e^{-\Gamma(t_0)}}{2}\right)^N.
 \end{equation}

 The $N$-dependent decay of  $\mathcal{F}(t_0)$ and the exponential decay of
    $\langle {\hat{J}^{(0)}}_{x}\rangle
$ reflect the
  rapid collapse of the  entanglement as  $N$ increases (See
  Fig.~2).
\begin{figure}[h]
\addtolength{\belowcaptionskip}{-0.6
cm}\addtolength{\abovecaptionskip}{-0.6 cm}
\begin{center}
\leavevmode {\includegraphics[width=4in,height=2.8in]{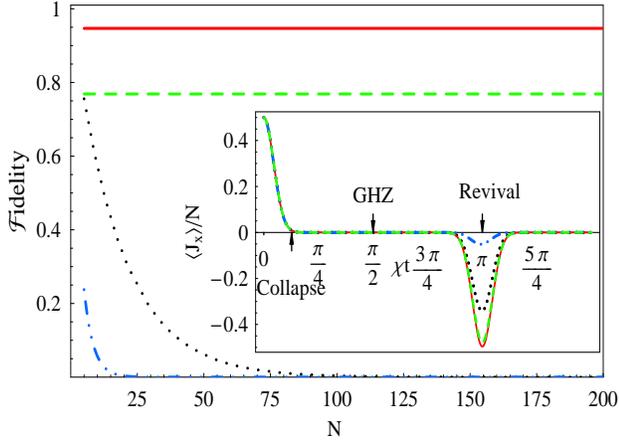}}
\end{center}
\caption{
(Color online)
Fidelity of GHZ generation vs. $N$  with and without gap
protection. We assume a noise with step-like spectral density with
 amplitude $f$ and cutoff frequency $\omega _{c}=\chi$ .
The dashed green and solid red lines are  for a protected system, $
\lambda=50$, with $f=0.6 \chi$
 and $f=0.1 \chi$ respectively. The dot-dashed blue and black dotted lines are for  an
unprotected system, $\lambda =0$, and  same decoherence parameters.
In the inset we show   $\langle {\hat{J}^{(0)}}_{x}(t) \rangle /N$
for $N=50 $. }
\end{figure}
Consider now the same generation
 process within the MPM. The latter can be realized by replacing
$\hat{H}_{z} \to \hat{H}_{prot} + \hat{H}_{z}$. In the slow noise limit
 the reduced density matrix evolves as Eq. (\ref{pde}) with the phase  $\omega_0(M-\tilde{M})$
 replaced by $\chi(M^2-\tilde{M}^2)$.
 For the initially polarized  state  the fidelity  is given by:
\begin{equation}
\addtolength{\belowdisplayskip}{-0.3
 cm}\addtolength{\abovedisplayskip}{-0.3 cm}
\mathcal{F}(t_0)=\frac{1}{\sqrt{1+\Gamma(t_0)}}\label{fid}.
 \end{equation}
The independence of $\mathcal{F}(t_0)$  on  $N$, and the
N-times reduced decay rate of  $\langle {\hat{J}^{(0)}}%
_{x}\rangle $  demonstrate  the  usefulness of MPM
 to generate  a large number of entangled particles.

Let us now discuss   physical implementations of the MPM. In Ref. \cite{Unanyan}
it has been shown that the Hamiltonian $\hat{H}_{prot} + \hat{H}_{z}$
 can be implemented  by using the collective vibrational motion of the ions
in a linear trap driven by illuminating them with a laser field. If
the  detuning of the laser, $\delta$, from  the internal transition
frequency is  large compared to the linewidth of the resonance but
sufficiently different from the ion vibrational frequency,  then the
dominant processes are two-photon transitions which lead to
simultaneous excitations of pairs of ions and thus  to an effective
Hamiltonian of the form:
$\hat{H}_{eff}=\chi(\hat{J}^{(0)2}-\hat{J}_z^{(0)2})+ \beta
 \hat{J}_z^{(0)}$. The cost of implementing $\hat{H}_{eff}$ instead of   $\hat{H}_{z}$
 is the additional echo techniques required for making the former insensitive to  heating of the vibrational
 motion of the ions: while  $\hat{H}_{z}$ is naturally insensitive to thermal vibrations
 due to the fact that it is generated by  bichromatic beams \cite{Sorensen} ( which eliminate  any dependence on
 vibrational quantum numbers due to the destructive interference between the transition paths)  $\hat{H}_{eff}$
 lacks  this insensitiveness as it is generated by monochromatic beams.
On the other hand, the
implementation of $\hat{H}_{eff}$
has the advantage in comparison with
$\hat{H}_{z}$ that the MPM
 protects the system from local perturbations
caused
by non-ideal
 conditions, such as   differences in the  Rabi frequency experienced by the
string of ions, or by  external disturbances, such as magnetic field
noise.
It should be noted, however, that the system is not protected by the MPM
against
spontaneous emission and
global perturbations.

{\it MPM with local interactions:} Up to now we have explored only
the generation of an MPM
  via isotropic long-range interactions.  In practice, however,
it is desirable to have a similar kind of protection  generated by
systems with short range interactions such as those provided by
 cold atoms in optical lattices. We now show how an MPM can be created in lattice  systems and can be
 used  to robustly  generate
$N$-particle GHZ states in these systems.
 We consider ultracold bosonic atoms with two
relevant internal states confined in  an
 optical lattice with unity filling
  deep in
the Mott insulator regime. Such systems is described by an effective
spin XXZ Hamiltonian \cite {Duan}: \begin{eqnarray}
\addtolength{\belowdisplayskip}{-0.2
 cm}\addtolength{\abovedisplayskip}{-0.2 cm}
\hat{H}_{lat}=\hat{H}_{H}+\hat{H}_{I}=-{\bar{\lambda}}\sum_{<i,j>,\alpha }\hat{%
\sigma}_{i}^{\alpha }\hat{\sigma}_{j}^{\alpha }-{\bar{\chi}}\sum_{<i,j>}\hat{%
\sigma}_{i}^{z}\hat{\sigma}_{j}^{z}.  \label{EQNBHH}
\end{eqnarray}Here we have identified the two possible states at each site with
the effective spin index $\sigma =\uparrow ,\downarrow $. The sum of
$\langle i,j\rangle $ is over nearest neighbors. In Eq.
(\ref{EQNBHH})
the coefficients are $\bar{\lambda}=\tau^{2}/U_{\uparrow \downarrow }$ and $\bar{\chi}%
=\tau^{2}(U_{\uparrow \uparrow }+U_{\downarrow \downarrow
}-2U_{\uparrow \downarrow })/U^{2}$, where $U$ is the mean of the three different spin  dependent on-site interactions energies
$U_{\uparrow \uparrow },U_{\downarrow \downarrow },U_{\uparrow
\downarrow }$, which we assumed only slightly different between each other, and $\tau$ is the tunneling energy between adjacent
sites which we assume spin independent. Both $U$ and $\tau$ are
functions of the lattice depth. For simplicity we restrict the
analysis to one dimensional systems and assume periodic boundary
conditions.

\begin{figure}
\addtolength{\belowcaptionskip}{-0.6
cm}\addtolength{\abovecaptionskip}{-0.6 cm}
\centering
\subfigure[]{\includegraphics[width=3.3 in,height=2.5 in]{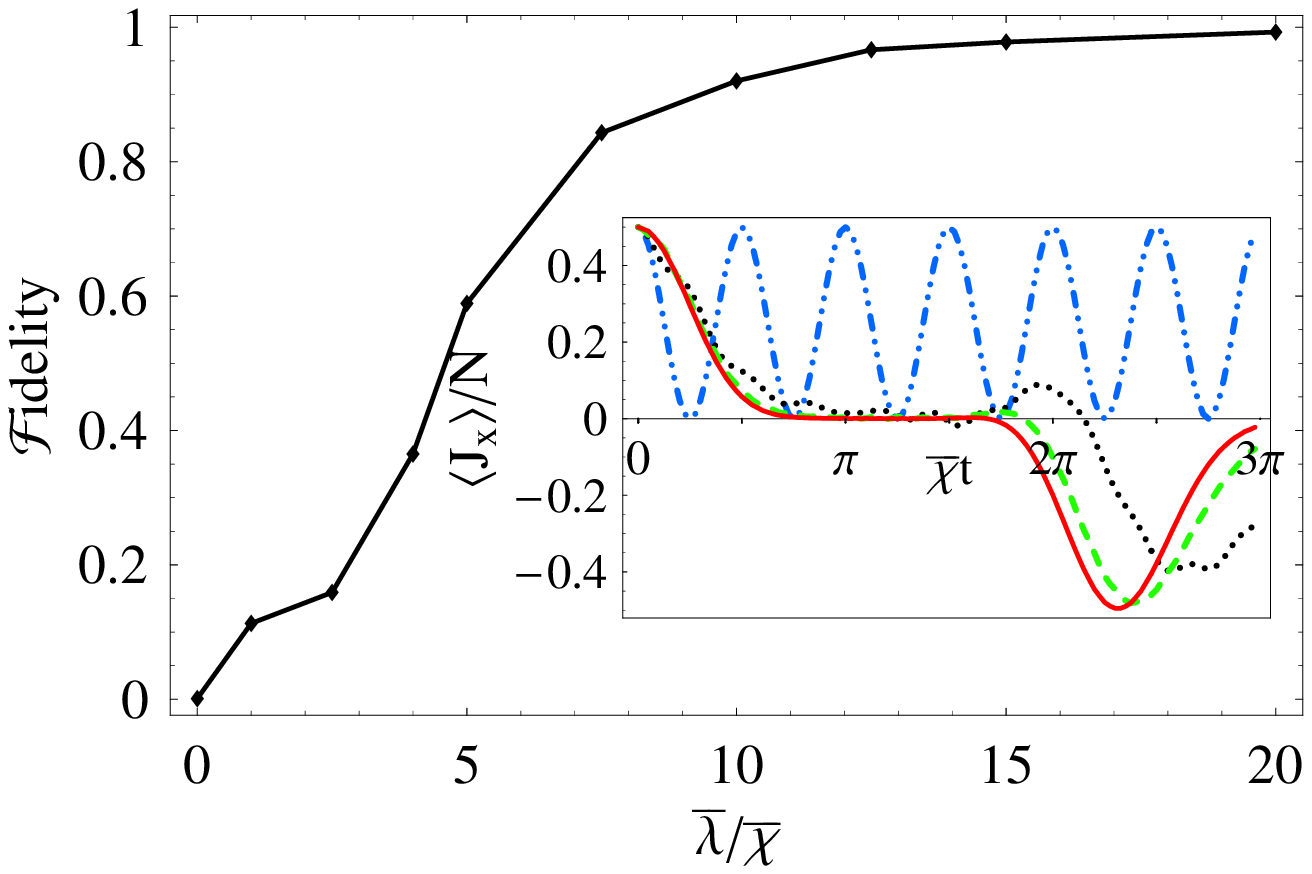}}\\
\subfigure[]{\includegraphics[width=3. in,height=2.2 in]{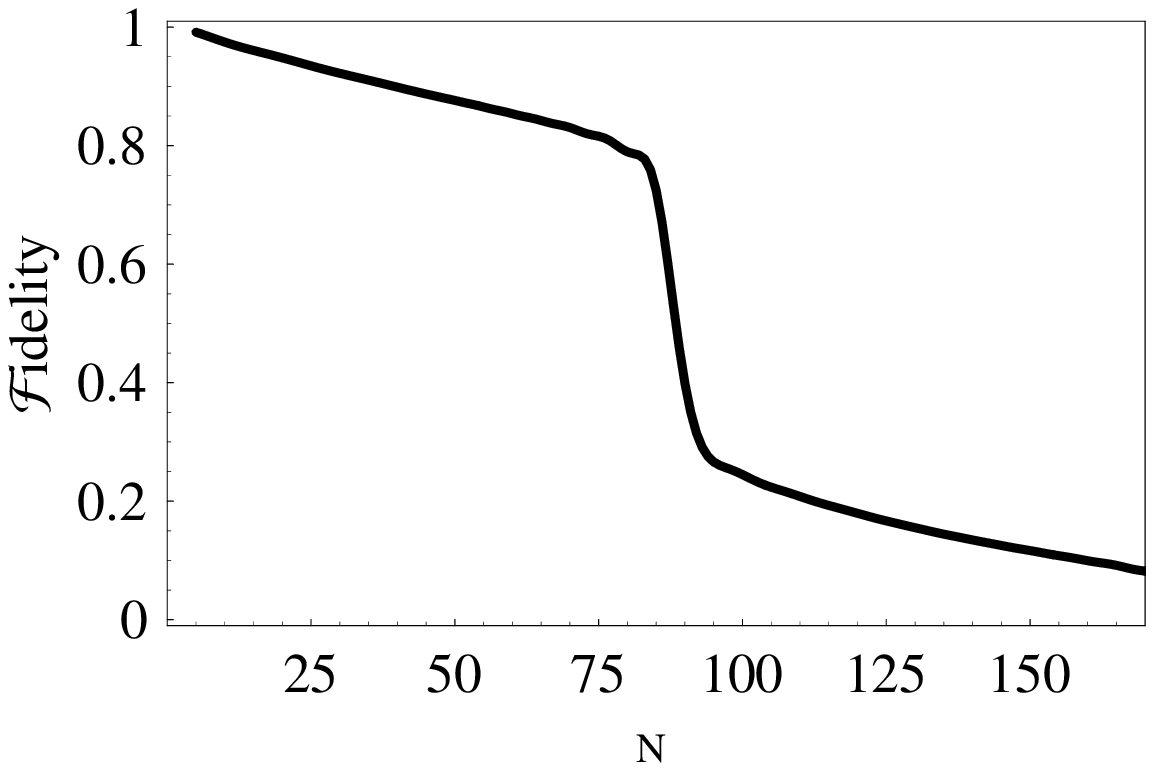}}
 \caption{(color online) a) Fidelity to generate a GHZ state vs $\bar{\lambda}/\bar{\chi}$. In
the inset we show $\langle{\hat{J}^{(0)}}_x(t)\rangle $. The blue
dot-dashed, dotted black,  dashed green and solid red correspond to
$\bar{\lambda} = 0, 5, 10, 20$ respectively. The plots are obtained
by  numerical evolution of Eq. (\ref{EQNBHH}) for  $N=10$.
b) In the presence of phase decoherence the fidelity of the GHZ state preparation in the lattice
is degraded as  $N$ grows
because in the lattice the gap decreases and the generation time
increases with  $N$. In this plot we used a
 system with
$\omega_c= \bar{\chi}$, $\bar{\lambda}=100 \bar{\chi}$ and
$\Gamma=0.01 \bar{\chi}$.  At $N=90$, $\Delta E_g=\omega_c$
and it explains  the drop of $\mathcal{F}$ for $N>90$.}
\end{figure}

 $\hat{H}_{H}$ is spherically symmetric and in terms of collective spin operators it can be written as
\begin{equation}
\addtolength{\belowdisplayskip}{-0.25
 cm}\addtolength{\abovedisplayskip}{-0.2 cm}
\hat{H}_{H}=-\frac{4\bar{\lambda}}{N}
\hat{J}^{(0)2}-\frac{4\bar{\lambda}}{N}\sum_{k>0, \alpha }{\hat{
J}_{\alpha}^{(k) }}{\hat{ J}_{\alpha}^{(-k) }}  \cos\left(\frac{2
\pi k }{N}\right).\label{bas}
\end{equation}All the  $N+1$ states with $J=N/2$ are degenerate and span  the
ground state of
 $\hat{H}_{H}$ thus form an MPM. $\hat{H}_{I}$ is not spherically symmetric but we can also write  it in terms of
collective operators as
\begin{equation}
\addtolength{\belowdisplayskip}{-0.25
 cm}\addtolength{\abovedisplayskip}{-0.2 cm}
\hat{H}_{I}=-\frac{4\bar{\chi}}{N}
\hat{J}^{(0)2}_{z}-\frac{4\bar{\chi}}{N}\sum_{k=1}^{N-1}\hat{
J}_{z}^{(k) } \hat{ J}_{z}^{(-k) } \cos\left(\frac{2 \pi k
}{N}\right).\label{bas}
\end{equation}
If the condition $\bar{\chi}\ll \bar{\lambda}$ is satisfied, which
is naturally the case for spin independent lattices,  the effect of
the Ising term
 can be
treated perturbatively.
 Assuming that at $t=0$ the initial
state is prepared within the $J=N/2$ manifold, a  perturbative analysis predicts
 that for times $t$ such that $\bar{\chi}t<\bar{\lambda}/\bar{\chi}$,
 $\hat{H}_{H}$ confines the dynamics to the MPM  and
transitions outside
 can be neglected. As a consequence, only
the projection of  $\hat{H}_{I}$  on  $\hat{H}_{H}$ is effective. As
$\mathcal {P}\hat{H}_{I}= {\chi }_{e} \hat{J}^{(0)2}_{z}+{\rm
const}$, with ${\chi }_{e}\equiv
\frac{4\bar{\chi}}{N-1}$\footnote{Here we used the relation
$\mathcal{P}_{k \neq 0}[\hat{ J}_{z}^{k} \hat{ J}_{z}^{-k
}]=-\frac{\hat{ J}_{z}^2}{N-1} +\frac{N^2}{4 (N-1)}$, with
$\mathcal{P}$ the projection into the $J=N/2$ subspace.}, $H_I$ acts
as a long range Hamiltonian.

In Fig.3a we show the fidelity to create a GHZ state at $\chi_e
t=\pi/2$ vs $\bar{\lambda}/\bar{\chi}$ and contrast the dynamical
evolution of $\langle{\hat{J}^{(0)}}_x\rangle$  for different
$\bar{\lambda}/\bar{\chi}$ ratios
 assuming  at time $t=0$ all the spins are polarized in the $x$ direction. If only the
Ising term is present, $\bar{\lambda}=0$, it induces local phase
fluctuations that leads to fast oscillations in
$\langle{\hat{J}^{(0)}}_x\rangle=N/2 \cos^2[2 \bar{\chi} t]$. On the
other hand, as the ratio $\bar{\lambda} / \bar{\chi}$ increases, the
isotropic interaction inhibits the fast oscillatory dynamics
 and instead
$\langle{\hat{J}^{(0)}}_x\rangle$ exhibits slow collapses and
revivals. For $\bar{\lambda} \gg \bar{\chi}$ the dynamics exactly
resembles the one induced by $\hat{H}_z$ and  at ${\chi }_{e}
t=\pi/2$ the initial coherent state is squeezed into a $GHZ$ state.

$\hat{H}_{H}$ also provides protection against phase decoherence. However, $\hat{H}_{H}$
is not as effective as  $\hat{H}_{prot}$ because
 the energy gap between the MPM and the excited states
 of $\hat{H}_H$ (Eq.\ref{bas}) vanishes in the thermodynamic limit  as $E_g \to \bar{\lambda}/N^2$ . This is a
  drawback  of  the short
range Hamiltonian for the purpose of  fully protecting the ground
states from long wave length excitations \footnote{ 1D systems are
the worst scenario as the gap vanishes
 as $N^{-2/D}$ with D the dimensionality}. Nevertheless,  the
many body interactions  can
 still  eliminate short-wavelength excitations   since in the large $N$ limit they remain separated by a finite
energy  gap, $8 \lambda$. In Fig.3b we quantify the protection
provided by the lattice Hamiltonian against population decay by
plotting $\mathcal{F} $ as a function of $N$ . As expected, an
abrupt drop of the fidelity occurs at the value of $N$ at which $
E_g =\omega_c$.

In {\it summary} we proposed and described a new method for
 the robust generation
of entangled states
and for their protection  against
decoherence and discussed
  its applicability for the generation of many-particle-GHZ-type states in ion traps and  optical lattices.
We emphasize  that, even though we have limited the discussion to
ensembles of spin $S= 1/2$ particles, the  MPM ideas can be
straightforwardly
 generalized to systems composed of  higher spin atoms.
Besides entanglement generation,  the MPM  might have also important
applications for  the implementation of good storage memories using
 for example nuclear spin ensembles in solid state \cite{Johnson} or
 photons\cite{Fleischhauer}.

This work was supported
by ITAMP, NSF (Career Program), AFOSR, the Sloan Foundation, and the David and
Lucille Packard Foundation.

\begin{spacing}{1.0}
\bibliographystyle{plain}

\end{spacing}
\end{document}